\documentclass[a4paper,fleqn,usenatbib]{mnras}

\usepackage[T1]{fontenc}
\usepackage{ae,aecompl}

\usepackage{graphicx}	
\usepackage{amsmath}	

\title[Configuration of the Martian dust rings]{Configuration of the Martian dust rings: Shapes, densities and size-distributions from direct integrations of particle trajectories}

\author[X. Liu and J. Schmidt]{
	Xiaodong Liu$^{(1, 2)}$
J\"urgen Schmidt$^{(2)}$\thanks{E-mail:jurgen.a.schmidt@oulu.fi}
\\
(1) School of Aeronautics and Astronautics, Sun Yat-sen University, China\\
(2) Space Physics and Astronomy Research Unit, University of Oulu, Finland
}

\date{Accepted XXX. Received YYY; in original form ZZZ}

\pubyear{2020}

\begin{document}
\label{firstpage}
\pagerange{\pageref{firstpage}--\pageref{lastpage}}
\maketitle

\begin{abstract}
It is expected since the early 1970s that tenuous dust rings are formed by grains ejected from the Martian moons Phobos and Deimos by impacts of hypervelocity interplanetary projectiles. In this paper, we perform direct numerical integrations of a large number of dust particles originating from Phobos and Deimos. In the numerical simulations, the most relevant forces acting on dust are included: Martian gravity with spherical harmonics up to 5th degree and 5th order, gravitational perturbations from the Sun, Phobos, and Deimos, solar radiation pressure, as well as the Poynting-Robertson drag. In order to obtain the ring configuration, simulation results of various grain sizes ranging from submicron to 100 microns are averaged over a specified initial mass distribution of ejecta. 
We find that for the Phobos ring grains smaller than about 2 microns are dominant; while the Deimos ring is dominated by dust in the size range of about 5-20 microns. The asymmetries, number densities and geometrical optical depths of the rings are quantified from simulations. The results are compared with the upper limits of the optical depth inferred from Hubble observations. We compare to previous work and discuss the uncertainties of the models.
\end{abstract}

\begin{keywords}
planets and satellites: rings -- Moon -- meteorites, meteors, meteoroids -- celestial mechanics
\end{keywords}



\section{Introduction} \label{section_introduction}
It is well-known that Saturn, Jupiter, Uranus and Neptune have ring systems. For Mars, no rings have been detected yet, but it was suggested by \citet{soter1971dust} that Mars should possess a ring system, which consists of dust grains originating from Phobos and Deimos. A number of studies contributed to the dynamical modeling of the suspected Martian dust rings, which are reviewed in several papers \citep{hamilton1996asymmetric, krivov1997martian, krivov2006search, zakharov2014dust}. Attempts to detect the rings are described by \citet{showalter2006deep}, \citet{zakharov2014dust}, \citet{showalter2017dust}. The Japan Aerospace Exploration Agency (JAXA) will launch the Martian Moons Exploration mission in 2024 with the Circum-Martian Dust Monitor \citep{kobayashi2018situ}, an instrument that will conduct in-situ dust measurements.

Recent studies of particle dynamics in the Martian dust rings were presented by \citet{makuch2005long} and \citet{krivov2006search}. In addition to solar radiation pressure and the Martian oblateness $J_2$, \citet{makuch2005long} included in their model the Poynting-Robertson (P-R) drag which is important for grains with long lifetime. They integrated the orbit-averaged equations of motion for dust grains to obtain the spatial structure of the Deimos torus. Furthermore, the Martian eccentricity ($e_\mathrm{Mars}$) combined with the three aforementioned perturbation forces was considered by \citet{krivov2006search}, who also integrated the orbit-averaged equations of motion. From their numerical simulations the configurations of the Phobos and Deimos dust rings were obtained and the ring optical depths were estimated. The effect of $e_\mathrm{Mars}$ on the dust dynamics was also discussed by \citet{showalter2006deep}.

In this paper we study the full dynamics of dust particles ejected from Phobos and Deimos in terms of direct numerical integrations of the equations of motion. Particle lifetimes and spatial configurations of number density are obtained for 12 different grain sizes ranging from $0.5\mu\mathrm{m}$ to $100\mu\mathrm{m}$. The steady state size distributions, dominant grain sizes, cumulative grain densities and geometric optical depths for the rings are derived by averaging over the initial mass distribution of ejecta. The major asymmetries of the rings are evaluated in a sun-fixed reference frame, allowing for quantitative predictions for future in-situ measurements by forthcoming missions. 

The paper is organized as follows. In Section \ref{section_production}, the mass production rates of dust from both source moons are estimated. The equations of motion and the numerical integrations are described in Section \ref{section_simulations}. The resulting lifetimes of various grain sizes are presented in Section \ref{section_lifetime}, and the effect of the planetary odd zonal harmonic coefficient $J_3$ is analyzed in Section \ref{section_J3}. In Section \ref{section_configuration}, we present the resulting spatial configuration of the Phobos and Deimos rings. A discussion of the model results and uncertainties, along with our conclusions, is given in Section \ref{section_comparison_uncertainties}.

\section{Dust production rates} \label{section_production}
The mass production rate of dust ejected from the surface of Phobos and Deimos by impacts of the hypervelocity interplanetary particles can be written as \citep{krivov2003impact}
\begin{equation} \label{eq:M+}
M^+= \alpha_\mathrm{G}\, F_\mathrm{imp}^{\infty}\, Y\,S \,,
\end{equation}
where $F_\mathrm{imp}^{\infty}$ is the mass flux of the interplanetary projectiles, $\alpha_\mathrm{G}$ accounts for the enhancement of the flux due to gravitational focusing by Mars, $Y$ is the yield (which denotes the ratio of the total mass of the ejected particles to the mass of the projectiles), and $S$ is the cross section of the source moon. The value of $\alpha_\mathrm{G}$ is calculated from Eq.~(14) in \citet{2006P&SS...54.1024S}. For Phobos we obtain $\alpha_\mathrm{G} \! \approx \! 1.003$, and a focused impact velocity $v_\mathrm{imp} \! = \! 15.30 \, \mathrm{km\,s^{-1}}$; for Deimos we have $\alpha_\mathrm{G} \! \approx \! 1.010$ and $v_\mathrm{imp} \! = \! 15.12 \, \mathrm{km\,s^{-1}}$. The gravitational focusing effect at Phobos and Deimos is low because of the small mass of Mars. The yield $Y$ is calculated with the empirical formula by \citet{2001Icar..154..391K}. For $F_\mathrm{imp}^{\infty}$ at Mars we adopt the same value as \citet{krivov2006search}, i.e.~$1 \times 10^{-15} \, \mathrm{kg\,m^{-2}\,s^{-1}}$, obtained from interplanetary dust models \citep{1985Icar...62..244G, divine1993five}, which is consistent by order of magnitude with the number from the Interplanetary Meteoroid Engineering Model (version 1.1) \citep{2011Mints, 2005AdSpR..35.1282D}. For a pure silicate surface of the moons, we obtain mass production rates $M^+$ of about $1.1 \times 10^{-1} \, \mathrm{g/s}$ for ejecta from Phobos and about $3.4 \times 10^{-2} \, \mathrm{g/s}$ for ejecta from Deimos. We note that there are large uncertainties in the estimate of the mass production rate (see Section \ref{section_comparison_uncertainties}).

\section{Numerical simulations} \label{section_simulations}
A well-tested numerical code \citep[see][]{liu2016dynamics, liu2018dust, liu2018comparison} is used to integrate the evolution of dust grains. Grains of sizes ranging from submicron to 100 microns are simulated: $0.5 \,\mathrm{\mu m}$, $1 \, \mathrm{\mu m}$, $2 \, \mathrm{\mu m}$, $5 \, \mathrm{\mu m}$, $10 \, \mathrm{\mu m}$, $15 \, \mathrm{\mu m}$, $20 \, \mathrm{\mu m}$, $25 \, \mathrm{\mu m}$, $30 \, \mathrm{\mu m}$, $40 \, \mathrm{\mu m}$, $60 \, \mathrm{\mu m}$, and $100 \, \mathrm{\mu m}$, including the effects of Martian gravity, gravitational perturbations from the Sun, Phobos, and Deimos, solar radiation pressure, as well as P-R drag. 

The shapes of Jupiter and Saturn are both nearly hemispherically and axially symmetric, which is not the case for Mars. Thus, in our model the Martian gravity field up to 5th degree and 5th order is considered. We use silicate as the material for dust and adopt a bulk density of 2.37 $\mathrm {g \, cm^{-3}}$ \citep{makuch2005long, krivov2006search}.

The equations of motion for dust particles in the Martian system read
\begin{equation} \label{equ_dynamic_model}
\begin{split}
\ddot{\vec r} = &GM_\mathrm M\nabla  \Bigg[ \frac{1}{r}+\frac{1}{r}\sum_{l=1}^{l_\mathrm{max}}\sum_{m=0}^{l}\left(\frac{R_\mathrm M}{r}\right)^l \times \\
& P_{lm}(\sin\phi) (C_{lm}\cos m\lambda + S_{lm}\sin m\lambda) \Bigg] \\
& + GM_{\mathrm P}\left(\frac{\vec r_{\mathrm {dP}}}{r_{\mathrm {dP}}^3}-\frac{\vec r_{\mathrm P}}{r_{\mathrm P}^3}\right) + GM_{\mathrm D}\left(\frac{\vec r_{\mathrm {dD}}}{r_{\mathrm {dD}}^3}-\frac{\vec r_{\mathrm D}}{r_{\mathrm D}^3}\right) \\
& + GM_\mathrm{S}\left(\frac{\vec r_\mathrm{dS}}{r_\mathrm{dS}^3}-\frac{\vec r_\mathrm S}{r_\mathrm S^3}\right) \\
& + \frac{3Q_\mathrm SQ_\mathrm {pr}\mathrm{AU}^2}{4r_\mathrm {Sd}^2\rho_\mathrm gr_\mathrm gc}\left\{\left[1-\frac{\dot r_\mathrm {Sd}}{c}\right]\hat{\vec r}_\mathrm {Sd} - \frac{\dot{\vec r}_\mathrm {Sd}}{c}\right\} \,.
\end{split}
\end{equation}
Here $G$ is the gravitational constant, $M_\mathrm M$ the mass of Mars, and $R_\mathrm M$ the Martian reference radius. We denote with $P_{lm}$ the associated Legendre functions of degree $l$ and order $m$, where $l_\mathrm{max}=5$ in this work and $C_{lm}$ and $S_{lm}$ are the spherical harmonics of the Martian gravity field. Further, $\phi$ and $\lambda$ are the latitude and longitude in the Martian body-fixed frame, respectively, $M_{\mathrm P}$ ($M_{\mathrm D}$) is the mass of Phobos (Deimos), $\vec r_{\mathrm {dP}}$ ($\vec r_{\mathrm {dD}}$) is the vector from the dust particle to Phobos (Deimos), and $\vec r_{\mathrm P}$ ($\vec r_{\mathrm D}$) is the vector of the Phobos (Deimos) position. Similarly, $M_{\mathrm S}$ is the mass of the Sun, $\vec r_{\mathrm {dS}}$ is the vector from the dust particle to the Sun, and $\vec r_{\mathrm S}$ is the radius vector of the Sun. The symbol AU denotes the astronomical unit, $Q_\mathrm S$ is the solar radiation energy flux at one AU, $Q_\mathrm{pr}$ is the solar radiation pressure efficiency, $\vec r_{\mathrm {Sd}} = -\vec r_{\mathrm {dS}}$, $\rho_\mathrm g$ is the bulk density of the grain, $r_\mathrm g$ is the grain radius, and $c$ is the light speed.

The values of the gravity spherical harmonics are taken from the Mars gravity model MRO120D \citep{konopliv2016improved}. The formulas for solar radiation pressure and the P-R drag are taken from \citet{burns1979radiation}. The size-dependent values of the solar radiation pressure efficiency $Q_\mathrm{pr}$ are computed from Mie theory \citep{mishchenko1999bidirectional,mishchenko2002scattering} for spherical grains (see Fig.~3 in \citet{liu2018dust}), using the optical constants for silicate grains from \citet{mukai1989cometary}. Due to the small size of Phobos ($R_\mathrm{P} \! \approx \!$ 11 km) and Deimos ($R_\mathrm{D} \! \approx \!$ 6 km), the dust ejection velocity is higher than the escape velocity but much lower than the orbital velocity of the moons \citep{horanyi1990toward, horanyi1991dynamics}. Therefore, it is a very good approximation to start grains directly from the orbits of Phobos and Deimos. It is known that the dynamical behavior of dust particles strongly depends on the solar longitude (Martian season) at the launch time \citep{hamilton1996asymmetric, krivov1997martian, makuch2005long, krivov2006search}. Thus, in our simulations 100 particles per grain size are launched with uniformly distributed initial mean anomalies of the Martian orbit around the Sun \citep{krivov1997martian, showalter2006deep}.

The motions of dust grains are simulated until they hit Phobos, Deimos, Mars, they escape from the Martian system, or for a maximum of 100,000 years. In order to save computation cost, we also stop the simulation when the fraction of grains remaining in orbit for certain sizes are less than 5\% (the integrations for the 100 particles per grain size run in parallel on a computer cluster provided by the Finnish CSC -- IT Center for Science). When we check for collisions of the dust particles with a given target (Phobos, Deimos, and Mars), at each time step of the integration we calculate the distance between the particle and the center of the target. If this distance is smaller than the target's radius, an impact occurs. When the grain is close to the target, in order to avoid overlooking the impact due to discrete time steps of the numerical integrators, a cubic Hermite interpolation is adopted to calculate the minimum distance between the particle and the target's center approximately (see \citet{chambers1999hybrid} and \citet{liu2016dynamics}). In our simulations, we store the slowly changing orbital elements including semi-major axis, eccentricity, inclination, argument of pericenter, longitude of ascending node, and true anomaly. Generally, we store 10 sets of orbital elements for one orbital period. The orbital segment between two consecutively stored sets of osculating elements is approximately considered as Keplerian. For denser output, each of these segments is further divided into intervals that are equidistant in time. Since dust particles are produced and absorbed continually (by hitting sinks or escape from the system), each discrete point corresponds to one particle in the steady-state ring configuration (see details in Sections 3.4 and 4 in \citet{liu2016dynamics} as well as Section 4 in \citet{liu2018dust}).

\section{Particle Lifetimes} \label{section_lifetime}
\begin{figure}
\centering 
\noindent\includegraphics[width=0.45\textwidth]{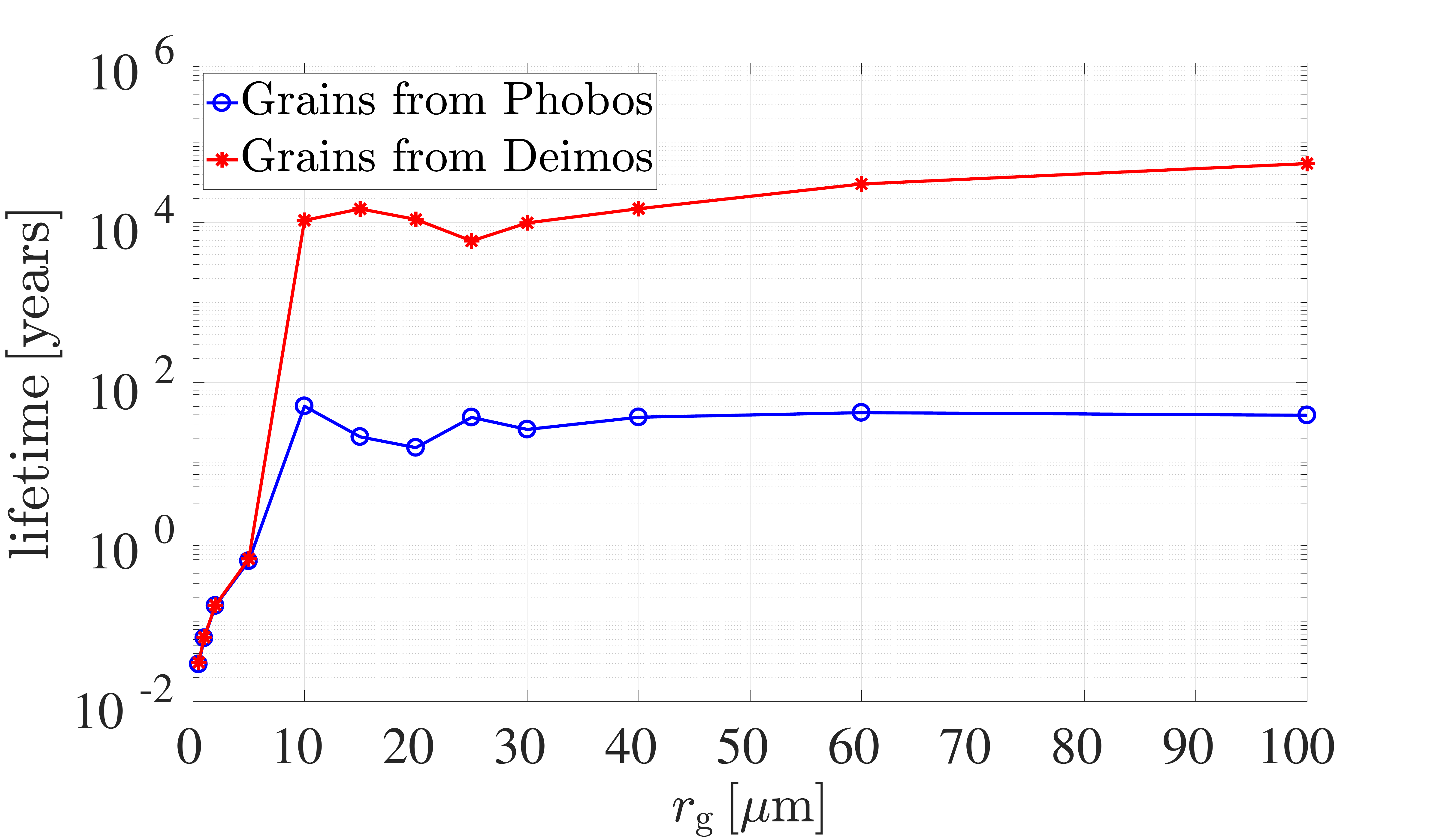}
\caption{Lifetimes for particles of various sizes originating from the Martian moons Phobos and Deimos.}
\label{fig:Lifetime}
\end{figure}

The lifetimes for dust from Phobos and Deimos derived from our simulations are shown in Figure ~\ref{fig:Lifetime}. Overall, we confirm the picture drawn from previous work (see table 1 in \citet{krivov2006search} and references given there): Grains larger than about $5\mu\mathrm{m}$ to $10\mu\mathrm{m}$ have lifetimes on the order of 10,000 years if they come from Deimos and a few tens of years if they are ejected from Phobos. For grain sizes smaller than 5 $\mathrm{\mu m}$ the lifetimes drop to a value of months. 

To explain this jump in lifetime, a conserved ``Hamiltonian" for the orbit-averaged system is used, which reads \citep{1996Icar..123..503H} 
\begin{equation}
\begin{split}
\label{eq:Hamiltonian}
\mathcal H (e, \phi_\odot) = & \sqrt{1-e^2} + Ce\cos\phi_\odot + \frac{1}{2}Ae^2\left[1+5\cos(2\phi_\odot)\right] \\
                             & + \frac{W}{3(1-e^2)^{3/2}} \ .
\end{split}
\end{equation}
Here $e$ is the eccentricity, $\phi_\odot$ is the solar angle, denoting the angle between the Sun and the grain's pericenter as seen from the planet, and $C$, $A$, and $W$ are parameters labeling the relative strengths of solar radiation pressure, solar tidal force, and the Martian $J_2$, respectively (see \citet{1996Icar..123..503H} for details). For this analysis, the perturbations from solar radiation pressure, the solar tidal force and $J_2$ are included, but other perturbations are ignored. Besides, the inclinations and eccentricities of Mars and source moons are neglected, and the grains have zero inclination. 

\begin{figure}
\centering 
\noindent\includegraphics[width=0.4\textwidth]{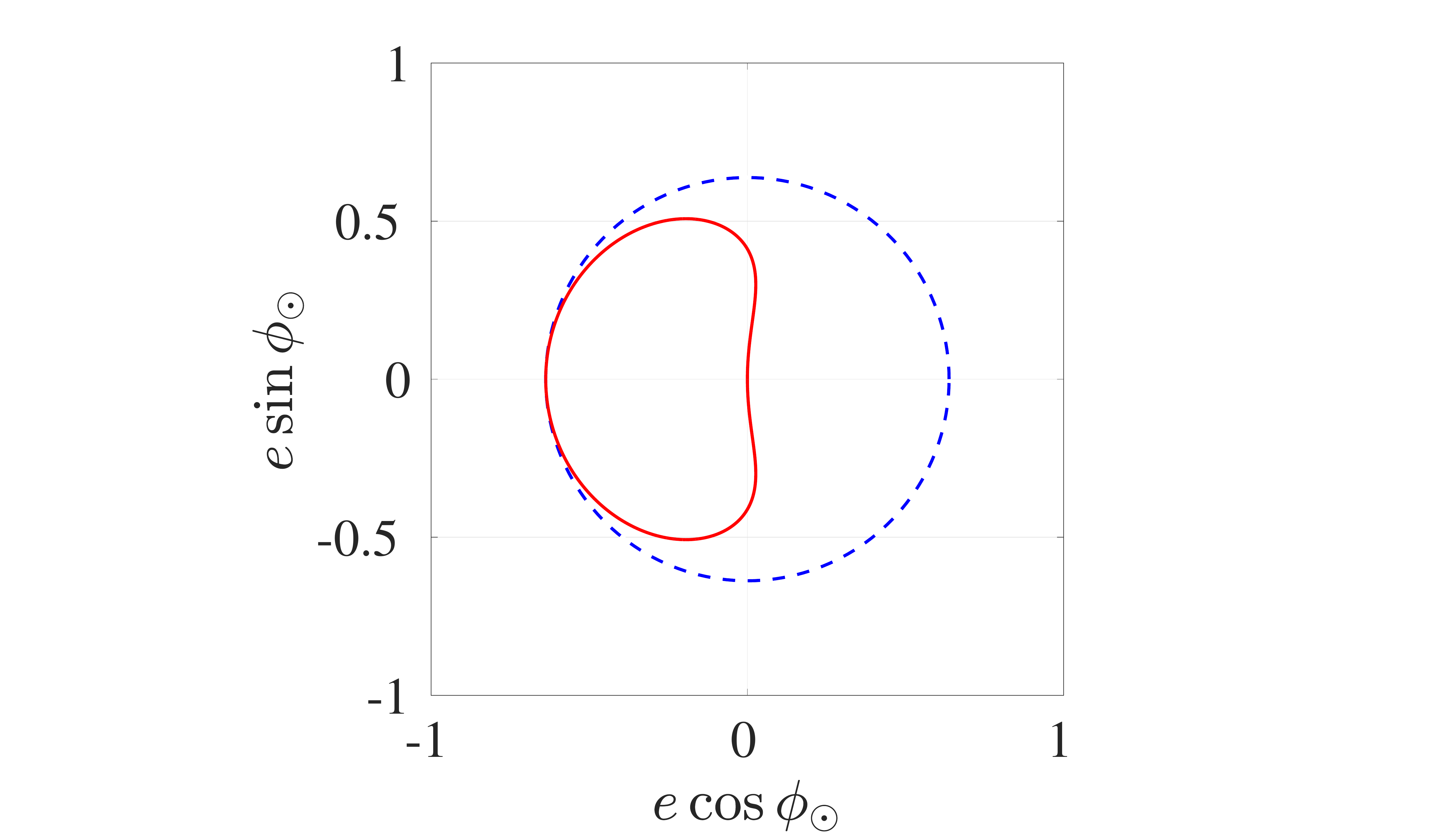}
\caption{Phase portrait for 10.7 $\mathrm{\mu m}$ particles from Phobos. The blue dashed circle denotes $e_\mathrm{impact}$ = 0.638 for Phobos grains.}
\label{fig:phase_phobos_impact}
\end{figure}

\begin{figure}
\centering 
\noindent\includegraphics[width=0.3\textwidth,angle=-90]{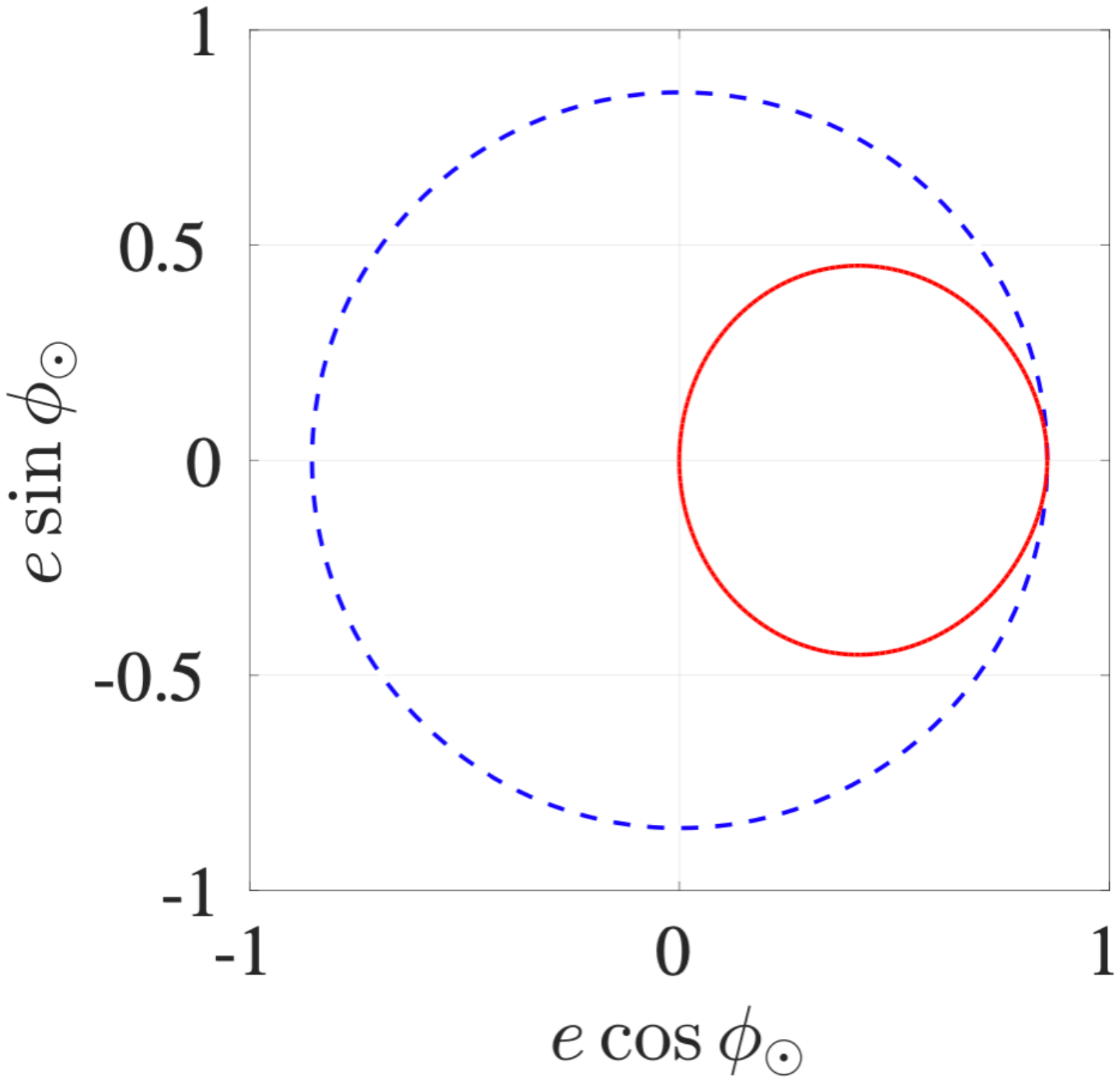}
	\caption{Phase portrait for 6.0 $\mathrm{\mu m}$ particles from Deimos. The blue dashed circle denotes $e_\mathrm{impact}$ = 0.855 for Deimos grains.}
\label{fig:phase_deimos_bifu_impact}
\end{figure}

Due to the low grain ejection velocity, which is negligible compared to the orbital velocity of the moons, the particles start their evolution with $e_0 \! \approx \! 0$. 
Since the Hamiltonian $\mathcal H$ is conserved, the orbital evolution of dust particles follows in the phase plane of the Hamiltonian a curve with the value at the starting point $\mathcal H (e=0, \phi_\odot)$. Typical phase portraits, obtained as a contour plot from Eq.~\ref{eq:Hamiltonian}, are shown in Figs.~\ref{fig:phase_phobos_impact} and \ref{fig:phase_deimos_bifu_impact}.

The value of eccentricity for which the grains' pericenter distance is equal to the Martian radius reads
\begin{equation}\label{eq:e_impact}
e_\mathrm{impact} = 1-\frac{R_\mathrm{M}}{a_\mathrm{moon}} \,,
\end{equation}
where $a_\mathrm{moon}$ is the semi-major axis of the source moon. The value of $e_\mathrm{impact}$ is of 0.638 for Phobos grains and of 0.855 for Deimos grains \citep{hamilton1996asymmetric}. With $r_\mathrm{g}^\mathrm{impact}$ we denote the grain size for which the maximum eccentricity $e_\mathrm{max}$, attained in the course of the orbital evolution, reaches $e_\mathrm{impact}$ \citep{hamilton1996asymmetric}. Particles smaller than $r_\mathrm{g}^\mathrm{impact}$ will develop a lager $e_\mathrm{max}$, and thus will hit Mars shortly after ejection, while grains larger than $r_\mathrm{g}^\mathrm{impact}$ will stay in the circum-Martian space for a much longer time. Based on Eq.~\ref{eq:Hamiltonian}, $r_\mathrm{g}^\mathrm{impact}$ = 10.7 $\mathrm{\mu m}$ for grains from Phobos (Fig.~\ref{fig:phase_phobos_impact}). Our analytical value of $r_\mathrm{g}^\mathrm{impact}$ is different from the one given by \citet{hamilton1996asymmetric} because we use size-dependent values of $Q_\mathrm{pr}$ for silicate grains (see Section \ref{section_simulations}). From the full numerical simulations, the value of $r_\mathrm{g}^\mathrm{impact}$ is close to but a bit smaller than 10.7 $\mathrm{\mu m}$, explaining the jump in the lifetime between 5 and 10 $\mathrm{\mu m}$ for Phobos grains (Fig.~\ref{fig:Lifetime}). From Fig.~\ref{fig:phase_phobos_impact} it is also evident that in principle we expect in the Phobos ring mostly particles with a solar angle in the range $90^\circ < \phi_\odot < 270^\circ $. For grains lifted from Deimos the phase portrait is shown in Fig.~\ref{fig:phase_deimos_bifu_impact}. For Deimos grains $r_\mathrm{g}^\mathrm{impact}$ = 6.0 $\mathrm{\mu m}$. 

Particles larger than $r_\mathrm{g}^\mathrm{impact}$ are safe from rapid collision with Mars. Their lifetime is limited mainly by collisions with their source moon and by the slow reduction of their semi-major axis, which, ultimately, increases again their chance to hit Mars. 


\section{The effect of $J_3$} \label{section_J3}
The effect of $J_2$ ($\approx \! 1.96 \times 10^{-3}$) on the dynamics of Martian dust was well studied in previous papers \citep[e.g.~][]{hamilton1996asymmetric, krivov1995dynamics}. In this section, the effect of a higher degree term, i.e.~the $J_3$ gravitational coefficient is analyzed. The $J_3$ term reflects an asymmetry in the mass distributions between the northern and southern hemispheres of the planet, which causes variations of inclination and eccentricity \citep{roy1965astrodynamics,paskowitz2006design}
\begin{equation} \label{equ_dot_i_J3}
\left\langle \frac{\mathrm{d}i}{\mathrm{d}t} \right\rangle_{J_3} = \frac{3nJ_3R_\mathrm{M}^3}{2a^3(1-e^2)^3} e \cos i \left(1-\frac{5}{4}\sin^2i \right) \cos\omega
\end{equation}
\begin{equation} \label{equ_dot_e_J3}
\left\langle \frac{\mathrm{d}e}{\mathrm{d}t} \right\rangle_{J_3} = -\frac{3nJ_3R_\mathrm{M}^3}{2a^3(1-e^2)^3} \sin i \left(1-\frac{5}{4}\sin^2i \right) \cos\omega \ .
\end{equation}
Here $i$ is the inclination, $n$ the mean motion, $a$ the semi-major axis, and $\omega$ the argument of pericenter.

\begin{figure}
\centering 
\noindent\includegraphics[width=0.45\textwidth]{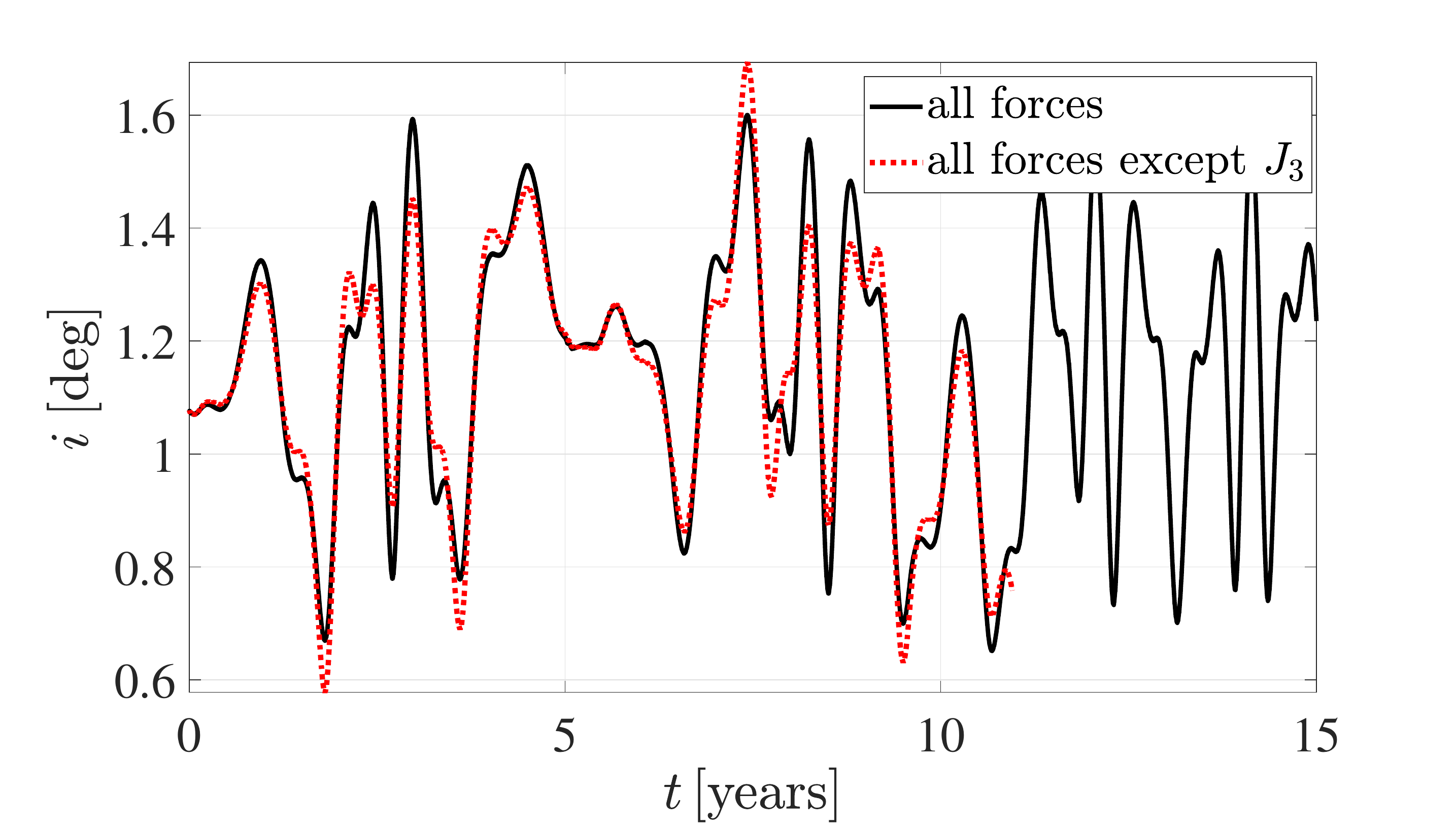}
\caption{Evolution of inclination for a 20 $\mu \mathrm{m}$ particle from Phobos. The black line denotes the inclination evolution with all perturbation forces (see Section \ref{section_simulations}). The red line corresponds to the case with all perturbation forces except $J_3$. Without $J_3$ the particle re-impacts on the source moon at an earlier time.}
\label{fig:dot_i_J3_phobos}
\end{figure}

For Jupiter and Saturn, because of their nearly hemispherically and axially symmetric shapes, the values of $J_3$ are almost zero ($J_3\approx -4.2 \times 10^{-8}$ for Jupiter \citep{iess2018measurement}, and $J_3 \! \approx \! 5.9 \times 10^{-8}$ for Saturn \citep{iess2019measurement}), and thus $J_3$ has negligible effect on the dynamics of particles in the Jovian and Saturnian rings. In contrast, Mars has a much larger value of $J_3$ ($\approx \! 3.15 \times 10^{-5}$), exceeding the value of $J_3$ for the Earth ($\approx -2.5 \times 10^{-6}$ \citep{pavlis2012development}), so that we might expect a noticeable effect on the dynamics of circum-Martian dust. From Eq.~\ref{equ_dot_i_J3}, $J_3$ could be important for the evolution of the inclination for eccentric orbits. 
In Fig.~\ref{fig:dot_i_J3_phobos} we show the evolution of a 20 $\mu \mathrm{m}$ particle from Phobos with and without the action of $J_3$. The maximal effect of $J_3$ on the inclination amounts to 10\%, roughly. The variations in inclination due to $J_3$, in turn, alter the effects of solar radiation and Martian $J_2$ on the evolution of semi-major axis and eccentricity. For a specific grain, the mildly altered dynamics can have a drastic effect on the lifetime (Fig.~\ref{fig:dot_i_J3_phobos}), while the overall, averaged effect on the lifetimes seems to remain small (see comparison to previous work in Section \ref{section_lifetime}).

Because of the factor $\sin i$ in Eq.~\ref{equ_dot_e_J3}, which is small due to the generally low orbital inclination, the direct effect of $J_3$ on eccentricity remains negligible compared to that of solar radiation pressure.

The $J_3$ effect decreases rapidly with increasing semi-major axis (Eqs.~\ref{equ_dot_i_J3} and \ref{equ_dot_e_J3}). Thus, it is more important for grains from Phobos than for those from Deimos. The $J_5$ gravitational coefficient has a similar, albeit much weaker effect on inclination. For particles from Phobos the strength of the perturbation induced by $J_5$ is roughly $J_5/J_3 \times \left(R_\mathrm{M}/a_\mathrm{Phobos}\right)^2 \! \approx \! 2\% $ of the perturbation induced by $J_3$.

\section{Particle size distribution and configuration of the rings} \label{section_configuration}
\begin{figure}
\centering 
\noindent\includegraphics[width=0.45\textwidth]{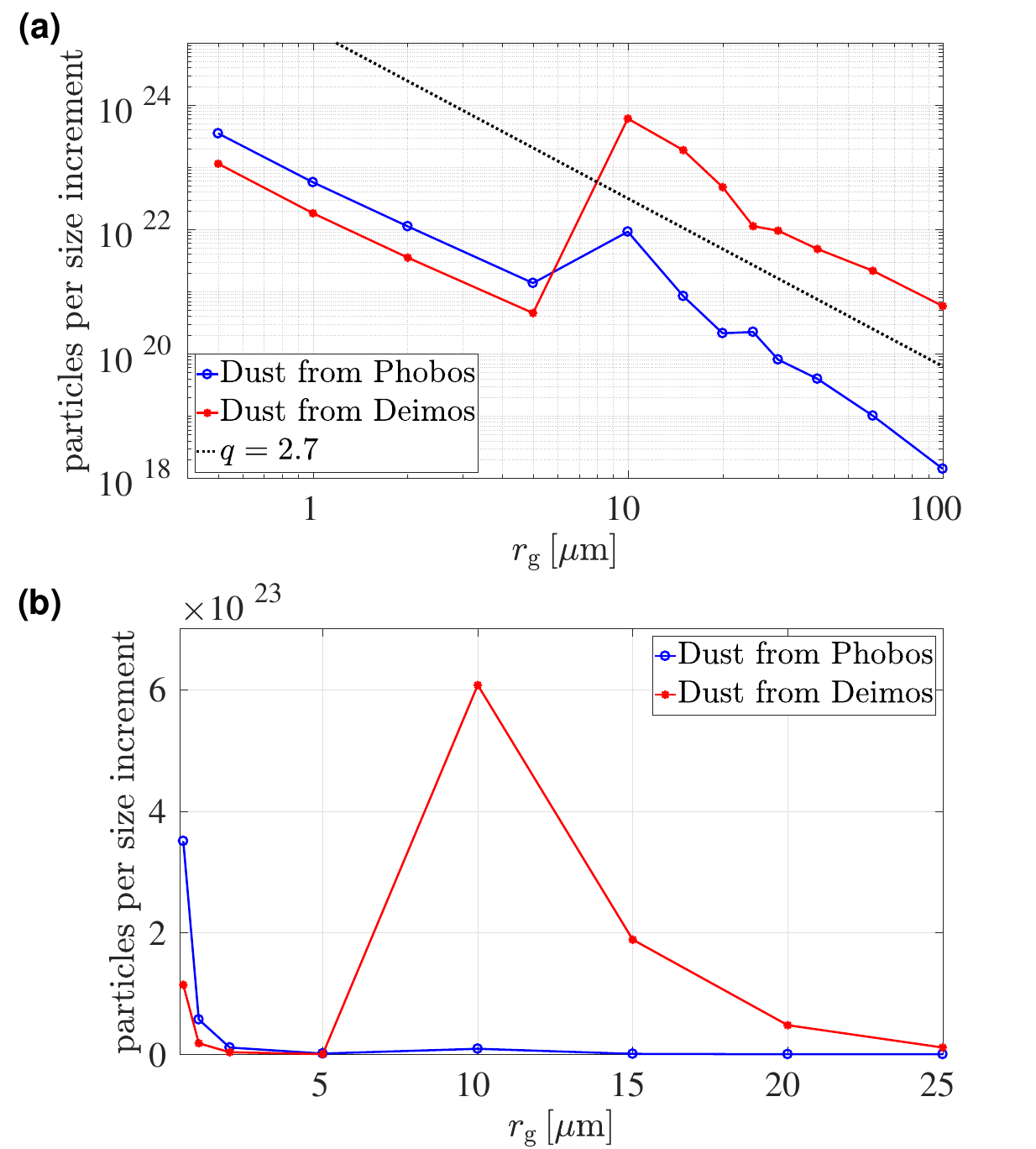}
	\caption{\textbf{(a)} Steady-state differential size distribution (logarithmic scale) for the Phobos and Deimos rings obtained from numerical simulations (Section \ref{section_simulations}) combined with initial mass distribution of ejecta Eq.~(\ref{equ_mass_distri}). A power law with slope $q$ = 2.7 is also shown for reference. \textbf{(b)} Same as \textbf{(a)}, but in linear scale for $r_\mathrm{g}<25\mu\mathrm{m}$.}
\label{fig:size_distri}
\end{figure}

The differential mass distribution of the ejected particles is assumed to follow a power law 
\begin{equation} \label{equ_mass_distri}
p(m) \propto m^{-(1+\alpha)} \,.
\end{equation}
For the slope $\alpha$ we adopt a value of $0.9$, derived from the in-situ measurements in the impact-generated lunar dust cloud \citep{Horanyi:2015faa} performed by the Lunar Dust Experiment onboard NASA's Lunar Atmosphere and Dust Environment Explorer mission. The mass distribution we normalize with the mass production rates given by Eq.~(\ref{eq:M+}). Averaging the results from the long-term simulations over the initial mass distribution (\ref{equ_mass_distri}) we obtain an estimate for the steady-state differential size distribution of grains in the Phobos and Deimos rings (Fig.~\ref{fig:size_distri}). Physically, this distribution arises from a folding of the steep size dependence of the initial grain production with the size dependent lifetime (Fig.~\ref{fig:Lifetime}).

\begin{figure}
\centering 
\noindent\includegraphics[width=0.4\textwidth]{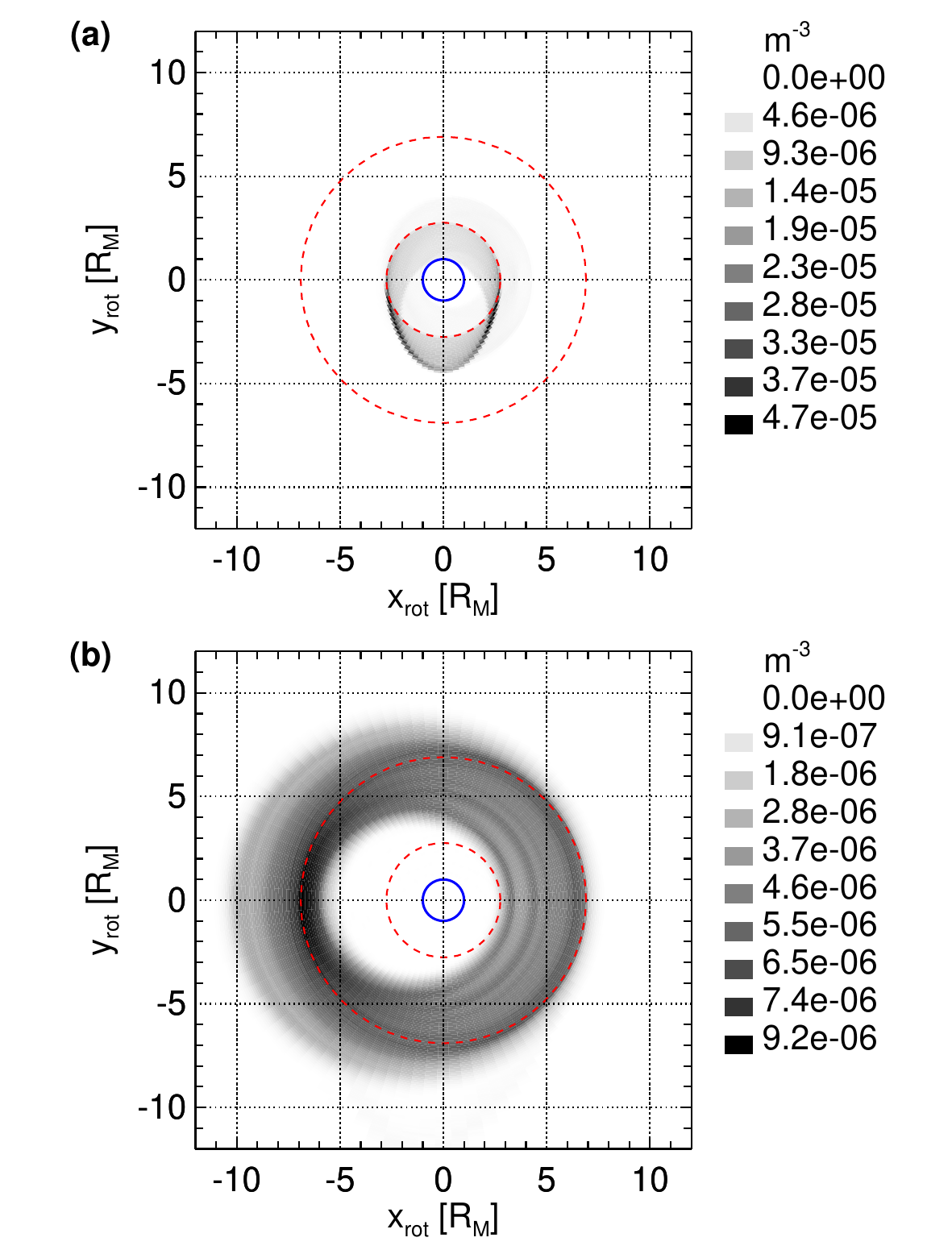}
\caption{\textbf{(a)} Grain number density in the Phobos ring projected onto the Martian equatorial plane, vertically averaged over $[-0.3, 0.3] \, R_\mathrm{M}$ in a frame that keeps a fixed orientation with respect to the Sun. The plot is obtained by averaging simulation results of particles of 12 grain sizes ranging from $0.5 \,\mathrm{\mu m}$ to $100 \,\mathrm{\mu m}$ (Section \ref{section_simulations}) over an initial mass distribution of ejecta with the differential slope $\alpha \! = \! 0.9$ (Eq.~\ref{equ_mass_distri}). See Fig.~\ref{fig:size_distri} for the steady-state differential size distribution for these ring particles. The positive $x_\mathrm{rot}$-axis points to the direction of the Sun. The blue line denotes the Martian radius, and the red dashed lines denote the orbits of Phobos (inner) and Deimos (outer). \textbf{(b)} Same as \textbf{(a)}, but for the Deimos ring, vertically averaged over $[-2.0, 2.0] \, R_\mathrm{M}$.}
\label{fig:normal_rotating_density}
\end{figure}

Since the lifetimes of larger grains from Phobos are not substantially longer than those of the small grains, we find that the Phobos ring is dominated by grains $\leq$ 2 $\mathrm{\mu m}$. A small, secondary peak around $10\mu\mathrm{m}$ is visible best in logarithmic scale (Fig.~\ref{fig:size_distri}). For the small grains that dominate the Phobos ring, solar radiation pressure is the most important perturbation force, which pushes nearly instantly the solar angle to $90^\circ$ \citep{hamilton1996asymmetric, 1996Icar..123..503H}. The subsequent rotation of the solar angle is induced by $J_2$ and by the orbital motion of Mars (see discussion around equation 5 of \citet{hamilton1996asymmetric}). Because of the small semi-major axis of Phobos, and thus of the grains lifted from its surface, the effect of $J_2$ dominates and the rotation is in anti-clockwise direction, assuming in principle values in the range $90^\circ < \phi_\odot < 270^\circ $ (see Fig.~\ref{fig:phase_phobos_impact}). But the rotation of the solar angle takes place on timescales that are longer than the lifetime of the grains. As a result, all grains develop their eccentricities with a sun-angle that remains close to $90^\circ$ and the Phobos ring appears shifted towards the negative $y$-axis (Fig.~\ref{fig:normal_rotating_density}). 

\begin{figure}
\centering 
\noindent\includegraphics[width=0.4\textwidth]{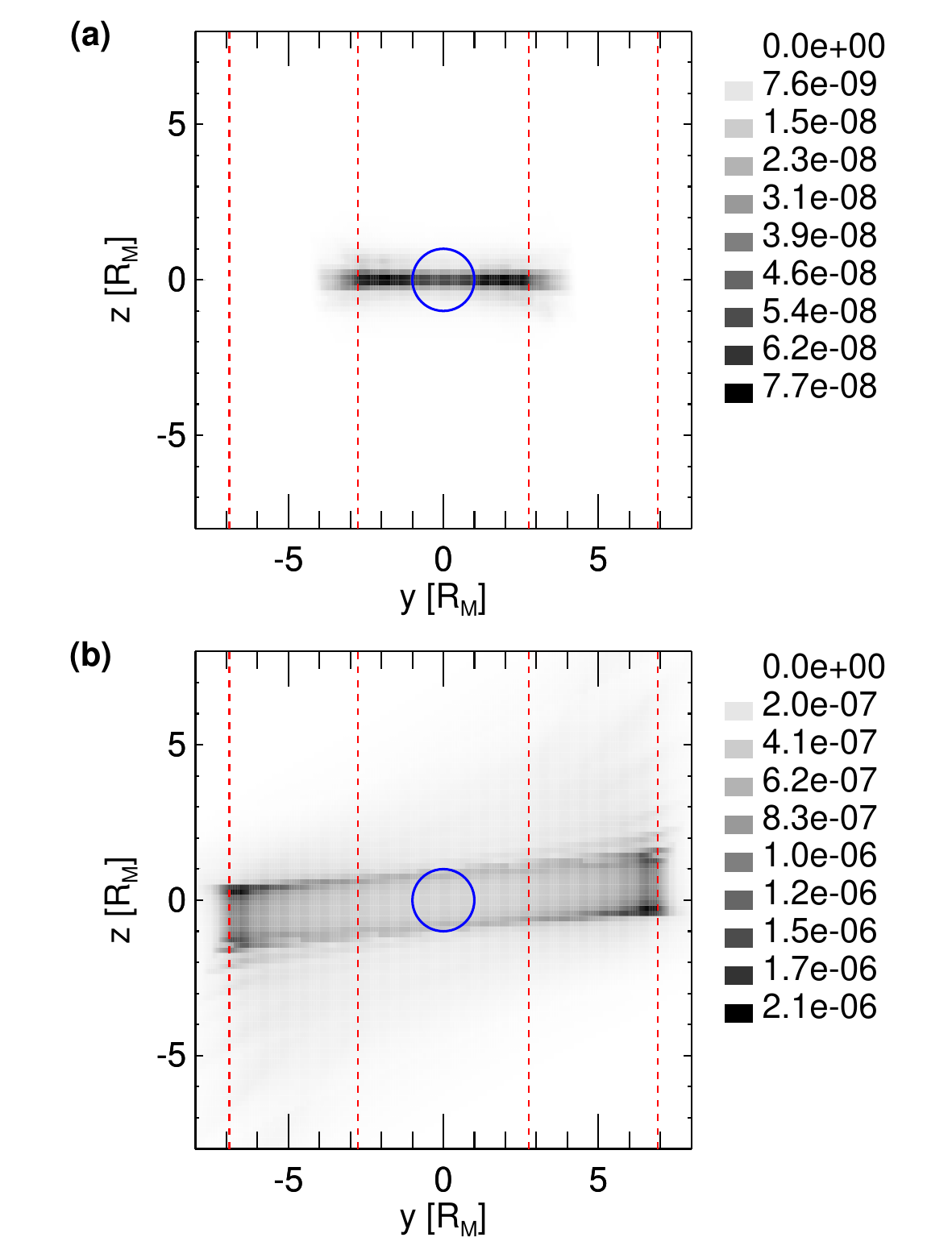}
\caption{\textbf{(a)} Geometrical optical depth of the Phobos ring, when viewed from the opposite direction of the Martian vernal equinox point. The plot is obtained by averaging simulation results of particles of 12 grain sizes ranging from $0.5 \,\mathrm{\mu m}$ to $100 \,\mathrm{\mu m}$ (Section \ref{section_simulations}) over an initial mass distribution of ejecta with the differential slope $\alpha \! = \! 0.9$ (Eq.~\ref{equ_mass_distri}). The $z$-axis is along the Martian spin axis. The $x$-axis (not shown in this plot) points to the Martian vernal equinox point, and the $y$-axis is perpendicular to the $x$-axis in the Martian equatorial plane. The blue line denotes the Martian radius, and the red dashed lines denote the orbital distances of Phobos (inner) and Deimos (outer). \textbf{(b)} Same as \textbf{(a)}, but for the Deimos ring.}
\label{fig:edge_yz_inertial}
\end{figure}

The Deimos ring, in contrast, is dominated by larger grains (Fig.~\ref{fig:size_distri}) in the size range of about 5-20 $\mathrm{\mu m}$, owing to the very large lifetimes of these particles (Fig.~\ref{fig:Lifetime}). This lifetime is longer than the period of the cycle in the evolution of the solar angle and the eccentricity (Fig.~\ref{fig:phase_deimos_bifu_impact}). For Deimos grains the rotation of the solar angle induced by the orbital motion of Mars dominates over the effect of $J_2$, so that the solar angle rotates clockwise \citep{hamilton1996asymmetric}. The maximum eccentricity is attained at $\phi_\odot \! = \! 0^\circ$. Averaging over many grains, lifted from Deimos uniformly over one Martian year, the Deimos ring appears shifted away from the Sun (Fig.~\ref{fig:normal_rotating_density}).

We find a peak number density (Fig.~\ref{fig:normal_rotating_density}) for grains $\geq$ 0.5 $\mu \mathrm{m}$ for the Phobos ring of about $4.7\times 10^{-5} \, \mathrm{m}^{-3}$ if vertically averaged over $[-0.3, 0.3] \, R_\mathrm{M}$ (corresponding roughly to the thickness of the ring, Fig.~\ref{fig:edge_yz_inertial}), and for the Deimos ring of about $9.2\times 10^{-6} \, \mathrm{m}^{-3}$ if vertically averaged over $[-2.0, 2.0] \, R_\mathrm{M}$. This allows us to estimate the number of dust impacts detected by an in-situ instrument on a spacecraft that traverses the rings. For a vertical crossing of the rings with a 1 $\mathrm{m}^2$ detector (as the sensitive area of the detector onboard JAXA's Martian Moons Exploration mission) one expects to record about 96 particles ($\geq$ 0.5 $\mu \mathrm{m}$) from the Phobos ring and about 125 particles from the Deimos ring.


The non-detection of the Martian rings in observations with the Hubble Space Telescope in 2001 \citep{showalter2006deep} placed upper limits on their edge-on brightness. For a geometric albedo of $0.07$, this translates into upper limits for the edge-on optical depth of $\sim2\times 10^{-6}$ for the Phobos ring and $\sim10^{-6}$ for the Deimos ring \citep{krivov2006search}. The edge-on geometric optical depths from our simulations are shown in Fig.~\ref{fig:edge_yz_inertial}. Here, the viewing direction is from the Martian vernal equinox point in the plane of the sky, i.e.\ the ascending node of the Martian orbital plane in the Martian equatorial plane. For this configuration we use a slope of $\alpha=0.9$ for the initial mass distribution (\ref{equ_mass_distri}) and we obtain an average edge-on geometric optical depth for the Phobos ring of about $3.5\times 10^{-8}$, and about $3.1\times 10^{-7}$ for the Deimos ring. If, alternatively, we use a value of $\alpha \! = \! 0.8$ \citep{krivov2003impact}, which is consistent with measurements of the dust cloud around the Galilean satellites by the Galileo Dust Detection System \citep{2000P&SS...48.1457K}, we get slightly larger values of about $3.7\times 10^{-8}$ for the Phobos ring and about $3.7\times 10^{-7}$ for the Deimos ring. In either case ($\alpha \! = \! 0.9$ or $\alpha \! = \! 0.8$) the average edge-on optical depth is lower than the upper limits inferred from the Hubble observations.

\section{Summary and Discussion} \label{section_comparison_uncertainties}
In this work we present a comprehensive numerical model for the evolution of dust particles ejected from the Martian moons Phobos and Deimos, with the goal to construct a steady state model for the configuration of the putative Martian dust rings. The new model ingredients are: (i) We perform direct numerical integrations of the equations of motion for a large number of particles, instead of integrating the orbit averaged evolution equations for the orbital elements. (ii) We use an array of 12 grain sizes, from submicron to 100 microns, to better resolve than previous studies the size dependent lifetimes. We present results as averages over an initial ejecta mass distribution. (iii) We include the Martian gravity field up to 5th degree and 5th order and account for the gravitational perturbations of Phobos and Deimos. (iv) We check for impacts of grains on the source moons and with Mars directly at (and between) the time steps of the integration, which allows for a more accurate evaluation of grain lifetimes than the probabilistic approach used in previous studies.

Our results are: \\
(1) We evaluate the lifetimes of grains with radii between $0.5\mu\mathrm{m}$ and $100\mu\mathrm{m}$, confirming results obtained in the literature. For grains from both source moons a jump in lifetime occurs between $5$ and $10$ microns. Smaller grains have lifetimes of months up to one (Earth) year. Larger grains from Phobos have lifetimes of tens of years while grains from Deimos remain in orbit for 10,000 years or more.\\
(2) The gravity perturbation induced by the Martian north-south asymmetry has in our simulations a small but noticeable effect on the orbital evolution of the grains from Phobos, in particular on the evolution of the inclination. For Deimos the effect is much smaller.\\
(3) Taking into account the initial mass distribution of ejecta, we derive the steady-state size distribution of dust particles in both ring components. The Phobos ring is dominated by grains smaller than a few microns. The Deimos ring is dominated by grains around $10$ microns in size. This is consistent with (but refines) previous results.\\
(4) Averaging over the initial mass distribution of ejecta and a large number of grains produced on the surfaces of the moons uniformly over one Martian year, we obtain a model for the steady state configuration of the rings. For the Deimos ring we confirm results from previous studies, in that the ring extends in the anti-sun direction, owing to an interplay of solar radiation pressure and the effect of Martian $J_2$ and the orbital motion of Mars. The ring has a thickness of about 4 Martian radii. For the Phobos ring we obtain the new result that the steady state ring should have a solar angle of $90^\circ$. This configuration arises from the dominance of small grains in this ring component and the correspondingly small lifetimes. The grains do not have time to perform a full cycle of the evolution of the eccentricity, and the solar angle stays close to its initial value of $90^\circ$. \\
(5) For a vertical traversal with a spacecraft, we estimate that a dust detector of $1\mathrm{m}^2$ area should record about 100 grains larger than $0.5\mu\mathrm{m}$ from either ring.\\
(6) We derive the edge-on geometric optical depth from our model, giving for the Phobos ring an estimate of $\tau\sim3.5\times 10^{-8}$ and about $\tau\sim3.1\times 10^{-7}$ for the Deimos ring, which is below the upper limits for the edge-on photometric optical depth inferred from observations of $\sim2\times 10^{-6}$ for the Phobos ring and $\sim10^{-6}$ for the Deimos ring.

Our model results are subject to fairly large uncertainties, as is the case for the models presented previously in the literature. The most uncertain parameter is the mass production rate of dust particles. On the one hand, the interplanetary projectile flux is still poorly constrained. On the other hand, not much is known about the surfaces properties of the source moons, which induces uncertainties in the ejecta yield, the bulk density of ejected grains, and in their dynamical response on solar radiation pressure and Poynting-Robertson drag. The values of the interplanetary flux and the ejecta yield affect the results linearly. The shape of the initial mass distribution, the bulk density of the particles, and their material have a size dependent effect, and their variation will affect the steady state size distribution, the number densities, and the optical depth of the rings in a non-linear manner. The precise error limits of the model results induced by the uncertainties in the parameters is difficult to assess, but it might easily amount to an uncertainty of an order of magnitude, or even more. For the initial distribution of ejecta masses we use a differential power-law with a slope of $\alpha=0.9$, as it was inferred by in-situ measurements in the lunar dust cloud. We also checked models with a slope of $\alpha=0.8$ (a value used in previous modelling of dust rings) and found that our main conclusions on the grain size distribution in the rings, the number densities, as well as their spatial configuration and the optical depths are robust. An additional complication might arise from the intermittency of the interplanetary flux \citep{Horanyi:2015faa}. For the small lifetimes of the grains that dominate the Phobos ring, this might result in a significant variability of the ring over months. Finally, we note that \citet{krivov2006search} pointed out the potential importance of grain-grain collisions as a sink, which we have not included in our modelling. Collisions might play a role especially for the Deimos ring, owing to the long particle lifetimes, leading to a depletion of the number density and optical depth. 

Although the Martian rings escaped detection so far, there is little or no doubt that the dust tori of Phobos and Deimos exist. The mechanism of quasi-continuous dust production in impacts of interplanetary meteoroids has been confirmed by measurements \citep{1999Natur.399..558K,Horanyi:2015faa}, as well as the formation of dust rings by this mechanism \citep{1999Sci...284.1146B,2004jpsm.book..241B,Hedman:2009kt}. The best chance to detect the rings might be in-situ measurements with a dust detector onboard a spacecraft, or, high-phase angle imaging from an orbiter when the spacecraft is in the shadow of Mars.

\section*{Acknowledgements}
This work was supported by the European Space Agency under the project Jovian Meteoroid Environment Model at the University of Oulu and by the Academy of Finland. We acknowledge CSC -- IT Center for Science for the allocation of computational resources.

\section*{Data availability}
The data underlying this article will be shared on reasonable request to the corresponding author.




\bibliographystyle{mnras}
\bibliography{Strings,AdditionalLit,PapersLit,lit,Ring2Galilean,trojan,mars}




\bsp	
\label{lastpage}
\end{document}